# What is a Planet?

**Steven Soter**

Department of Astrophysics, American Museum of Natural History,
Central Park West at 79th Street, New York, NY 10024, and
Center for Ancient Studies, New York University

soter@amnh.org

**Abstract**. A planet is an end product of disk accretion around a primary star or substar. I quantify this definition by the degree to which a body dominates the other masses that share its orbital zone. Theoretical and observational measures of dynamical dominance reveal gaps of four to five orders of magnitude separating the eight planets of our solar system from the populations of asteroids and comets. The proposed definition dispenses with upper and lower mass limits for a planet. It reflects the tendency of disk evolution in a mature system to produce a small number of relatively large bodies (planets) in non-intersecting or resonant orbits, which prevent collisions between them.

## 1. Introduction

Beginning in the 1990s astronomers discovered the Kuiper Belt, extrasolar planets, brown dwarfs, and free-floating planets. These discoveries have led to a re-examination of the conventional definition of a “planet” as a non-luminous body that orbits a star and is larger than an asteroid (see Stern & Levison 2002, Mohanty & Jayawardhana 2006, Basri & Brown 2006).

When Tombaugh discovered Pluto in 1930, astronomers welcomed it as the long sought “Planet X”, which would account for residual perturbations in the orbit of Neptune. In fact those perturbations proved to be illusory, and the discovery of Pluto was fortuitous. Observations later revealed that Pluto resembles neither the terrestrial nor the giant planets. Pluto is smaller than seven moons in the solar system, and its orbit crosses that of Neptune with a 3:2 mean motion resonance. For six decades Pluto remained a unique anomaly at the outer edge of the planetary system. Then, beginning in 1992, the discovery of other Kuiper Belt objects (KBOs) revealed that Pluto actually belongs to a vast population of icy trans-Neptunian objects (Luu & Jewitt 2002). This revelation challenged the conventional status of Pluto as a planet. The recent discovery of Eris (formerly known as UB313), a KBO larger than Pluto (Bertoldi et al. 2006), intensified the debate (Sheppard 2006).

A similar challenge accompanied the recognition of the asteroid belt. When Piazzi discovered Ceres in 1801, astronomers welcomed it as the missing planet that filled the gap between the orbits of Mars and Jupiter, according to Bode's Law (later shown to be specious). But in 1802 Olbers discovered Pallas, with practically the same semi-major axis as Ceres. Herschel recognized that both unresolved objects must be far smaller than planets and proposed naming them "asteroids". Olbers suggested that they were fragments of a disrupted planet. Two more asteroids were found by 1807, and for the next four decades astronomy textbooks listed all four bodies as planets, each with its own symbol. Between 1845 and 1851, the population of known asteroids increased to 15, and the continued planetary status of these small bodies became unwieldy. Astronomers then began to number all asteroids by their order of discovery, rather than by semi-major axis, as for planets (Hilton 2001). This marked the *de facto* acceptance of the asteroids as members of a population distinct from planets.

We now recognize that the solar system includes several distinct populations – the planets, satellites, asteroid belt, Kuiper Belt, Oort Cloud, etc. -- that reflect different pathways in the evolution of the solar nebula. The conventional list of "nine planets" -- four terrestrial planets, four giant planets, and Pluto – has lost any scientific rationale, and is now merely historical. If Pluto is included as a planet, we have no physical basis for excluding Eris, dozens of other large spherical KBOs, and Ceres. The term "planet" would then lose any taxonomic utility. But an important function of scientific nomenclature is to reflect natural relationships, not to obscure them.

Attempts to define "planet" in terms of upper and lower mass limits have not been satisfactory. An upper mass limit corresponding to the onset of deuterium fusion is complicated by the existence of some brown dwarfs in close orbits around stars (see Section 6). A lower mass limit to distinguish planets from smaller non-planets is also problematic.

Stern and Levison (2002) suggested a lower size limit for a planet based on the criterion of hydrostatic equilibrium. Any non-stellar body large enough for its gravity to dominate its shape would be a planet. Such a criterion, however, involves not only the size but also the density and compressive strength of the material. For example, the rocky asteroid Vesta (538 km diameter) is clearly non-spherical, while the icy satellite Mimas (395 km) looks round (Basri & Brown 2006). Also, how should one quantify the limiting shape that distinguishes a planet? In a population of small bodies spanning a continuum of sizes and shapes, does gravity dominate the shape of a body if the cross-section deviates from hydrostatic equilibrium by 10%, or by 1%? Nature provides no unoccupied gap between spheroidal and non-spheroidal shapes, so any boundary would be an arbitrary choice.

## 2. Dynamical Dominance in the Solar System

Nature does, however, provide a suitable criterion for planetary status based on a wide gap in a physically significant parameter – namely the measure of the extent to which a body dominates the other masses in its orbital zone. Stern and Levison (2002) remarked that some bodies in the solar system are dynamically important enough to have cleared out most of the neighboring planetesimals in a Hubble time, while lesser bodies, unable to do so, occupy transient unstable orbits, or are preserved in mean motion resonances or satellite orbits. Applying the techniques of Öpik (1976), they derived a parameter $\Lambda$ to quantify the extent to which a body scatters smaller masses out of its orbital zone in a Hubble time,

$$\Lambda = kM^2/P, \qquad [1]$$

where k is approximately constant and M and P are the scattering body's mass and orbital period, respectively. We note that $\Lambda = T/\tau$, where T is the Hubble time and $\tau \propto P/M^2$ is a characteristic timescale for scattering or ejection of small bodies from the vicinity of a body of mass M and period P (cf. Grazier et al. 1999a, Goldreich et al. 2004). A heliocentric body with $\Lambda > 1$ has cleared a substantial fraction of small bodies out of its orbital neighborhood.

Stern and Levison found a gap of five orders of magnitude in $\Lambda$ between the smallest terrestrial planets and the largest asteroids and KBOs (see Table 1). However, they did not take advantage of this large gap to define the term planet, as distinct from asteroids and comets. Rather they introduced the additional terms "überplanets" and "unterplanets" for bodies with $\Lambda > 1$ and $\Lambda < 1$, respectively.

Figure 1 plots mass M versus semi-major axis a for heliocentric bodies. The solid lines represent the observed boundaries of the gap in $\Lambda$, corresponding to its values for Mars and Pluto. The dashed line corresponds to $\Lambda = 1$. For constant $\Lambda$, Eq. [1] gives $M \propto a^{3/4}$. The objects above the gap are effectively solitary and completely dominate their orbital zones, while those below the gap live amid a swarm of comparable bodies. The region between these limits appears so far to be unoccupied in our solar system. While KBOs larger than Eris may yet be discovered, wide field surveys would probably have detected any objects larger than Mars on low inclination orbits within about 70 AU (Morbidelli et al. 2003)

Brown (2004) proposed a definition of "planet" based on the natural division of objects into solitary bodies and members of populations. A planet is "any body in the solar system that is more massive than the total mass of all of the other bodies in a similar orbit." For example, the planet Neptune has 8600 times the mass of Pluto, the largest body that crosses its orbit. Likewise, the planet Earth has $2 \times 10^8$ times the mass of the asteroid (1036) Ganymed, the largest body that crosses its orbit. In contrast, the asteroids and KBOs are members of populations with a shared orbital space, in which no member so dominates the others by mass. The two largest asteroids, Ceres and Pallas, differ in mass by a factor of about 4 (Kovacevic & Kuzmanoski 2005, Goffin 2001). Our solar system has no intermediate cases between solitary bodies (planets) and members of populations, defined in this way.

A modification of Brown's definition can link it explicitly to the dynamics of planet formation: A planet is a body that has swept up or scattered most of the mass from its orbital zone in the accretion disk around a central star or substar. In this paper I propose an observational criterion to quantify this definition.

The end product of secondary disk accretion is a small number of relatively large bodies (planets) in either non-intersecting or resonant orbits, which prevent collisions between them. Asteroids and comets, including KBOs, differ from planets in that they can collide with each other and with planets.

Planets, defined in this way, are few in number because the solar system provides insufficient dynamical room for many. Perturbations by the outer planets destabilize the potential orbits between them on time scales much shorter than the age of the solar system (Grazier et al. 1999b).

## 3. Definitions

(1) A "primary" body is a star or substar formed by core accretion from an interstellar cloud, not by secondary accretion from a disk.

(2). A "substar" is a body with less than 80 Jupiter masses, the lower limit for stellar hydrogen fusion.

(3) A "planet" is an end product of secondary accretion from a disk around a primary body.

(4) An “end product” of disk accretion is a body containing more than 100 times the mass of all other bodies that share its orbital zone.

(5) Two bodies share an “orbital zone” if their orbits cross a common radial distance from the primary, and their periods are non-resonant and differ by less than an order of magnitude.

To determine whether a body of mass M is an end product of disk accretion, let

$$\mu = M/m,$$

where m is the aggregate mass of all the other bodies that currently share its orbital zone. I refer to μ as the planetary discriminant. If $\mu > 100$ for any body orbiting a primary star or substar, then that body is by definition a planet.

This definition of planet applies only to mature systems, such as ours, in which accretion has run effectively to completion, and the major bodies can no longer undergo orbital migration. For younger evolving systems, where accretion is still important, the largest bodies are called “planetary embryos,” and the smaller bodies are “planetesimals.”

## 4. Census of Heliocentric Small Bodies in the Solar System

In order to estimate μ for various solar system bodies, one needs a census of objects in orbits that allow collisions. Three primary reservoirs supply most of the mass that continues to collide with solar system bodies. (1) The asteroid belt is a zone of rocky/metallic bodies and dormant comets in prograde orbits between Mars and Jupiter, mostly from 2 to 3.6 AU. (2) The Kuiper Belt is a toroidal region of comets in prograde orbits, between the orbit of Neptune and about 56 AU (Gladman 2005). (3) The Oort Cloud is a spherical region of comets that extends out to about $10^5$ AU.

These three primary reservoirs feed objects into five secondary populations, which are the proximate sources for collisions with the planets: near-Earth objects (NEOs), Mars-crossing objects (MCOs), Centaurs, short-period (SP) comets, and long-period (LP) comets. The asteroid belt supplies the NEOs and MCOs, which can collide with terrestrial planets. The Kuiper Belt supplies the Centaurs and SP comets. The Centaurs are comets orbiting between Jupiter and Neptune that usually cross the orbit of at least one giant planet (Tiscareno & Malhotra 2003), while the SP comets have periods $P < 200$ years and eccentricities that bring them well inside the orbit of Jupiter. The Oort Cloud feeds the entire planetary zone with LP comets having P > 200 years and isotropically distributed inclinations.

In defining μ, I chose for m the mass of the transient population of bodies currently within orbital range of collisions with each of the potential targets. This seems more realistic than invoking the mass of the entire asteroid belt for m with respect to the terrestrial planets, or the mass of the entire Oort Cloud for m with respect to all the planets. Only a minuscule fraction of the Oort Cloud mass would even be available for potential collisions with planets in a Hubble time. Another alternative would be to define m as the total mass expected to collide with each planet in a Hubble time, but such a calculation encounters too many unknowns. The choice for m of the present transient population masses at least provides a practical and observable discriminant for planets in our own system.

Table 2 lists the primary reservoirs and proximate populations of objects that can collide with planets, including their orbital semi-major axes a and perihelia q, the estimated number of objects with effective diameters D > 1 km, and their integrated mass.

**Primary Reservoirs**

***Asteroid Belt***. Tedesco and Desert (2002) estimated the total number of main belt asteroids with D > 1 km to be 1.2 x $10^6$. Krasinsky et al. (2002) calculated the mass of the asteroid belt at 6 x $10^{-4}$ $M_E$ (where $M_E$ = Earth mass).

***Kuiper Belt***. Bernstein *et al.* (2002) estimated the mass of the Kuiper Belt at 0.03 + 0.01 $M_E$. The masses and sizes of the comets remain poorly determined. I take an average effective comet radius of 5 km (Lamy et al. 2005) and a density 0.6 gm/$cm^3$ (A'Hearn et al. 2005), giving an average mass of about 3 x $10^{17}$ gm. The number of KBOs in Table 2 assumes this average comet mass.

***Oort Cloud***. Francis (2005) estimated that the Oort Cloud includes some 5 x $10^{11}$ comets, with a total mass of 2 to 40 $M_E$. I adopt 25 $M_E$, based on the average comet mass estimated here.

**Proximate Populations**

***Near-Earth objects*** (NEOs) are mostly asteroids with perihelia q < 1.3 AU. Stuart and Binzel (2004) estimated the population with D > 1 km to be 1090. Assuming the NEOs are a representative sampling by mass of the main asteroid belt, I estimate their aggregate mass at about 6 x $10^{-7}$ $M_E$.

***Mars-Crossing objects*** (MCOs) are asteroids that enter the heliocentric range 1.3 to 1.7 AU. Michel et al. (2000) estimate that they outnumber NEOs by a factor of about 35, so I take their aggregate mass to be about $2.1 \times 10^{-5}$ $M_E$.

***Centaurs***. Shepard et al. (2000) estimated the population and total mass of the Centaurs at about $10^7$ and $5 \times 10^{-4}$ $M_E$, respectively. Their total mass is comparable to that of the main asteroid belt.

***SP Comets.*** Levison et al. (2002) estimated the number of SP comets, the majority of which are dormant, at about $10^3$, so I take their aggregate mass to be about $5 \times 10^{-8}$ $M_E$. The population of SP comets is comparable to that of the NEOs, but they are distributed over a larger volume of orbital space, and to first order we neglect their contribution to the collision flux for the inner planets.

***LP Comets.*** At any given time, a small fraction of the Oort Cloud comets have highly eccentric (near parabolic) orbits with perihelia in the inner solar system. Although they spend most of the time more than $10^3$ AU from the Sun, these LP comets are the only members of the Oort Cloud with orbits that allow them to collide with the planets. To estimate $\mu$ for LP comets, I count only those LP comets that at any given time are within 50 AU of the Sun.

The average semi-major axis of the LP comets is about 30,000 AU (Wiegert & Tremaine 1999), corresponding to an orbital period of $5 \times 10^6$ yr. The Oort Cloud feeds LP comets into orbits with perihelia $q < 8$ AU at a rate of about 40 $yr^{-1}$ (Francis 2005). The perihelia of LP comets are distributed almost uniformly with heliocentric distance in this range, so the integrated number of such comets with $q < q_o$ is nearly proportional to $q_o$. Hence the flux of LP comets entering a sphere of given heliocentric radius r, per unit interval of orbital perihelion q, is about $\nu = 5$ $yr^{-1}$ $AU^{-1}$. The average number of LP comets with $r < r_o$ is then $N(r_o) = 2\nu\int t(q, r_o)dq$, where $t(q, r_o)$ is the time for a comet to fall from $r_o$ to its perihelion q, and the integral over q is from 0 to $r_o$. Approximating the orbits of LP comets as parabolas simplifies the calculation of $t(q, r_o)$.

The total number of LP comets in the inner solar system ($r_o < 3.6$ AU) at any time, potentially able to collide with terrestrial planets or asteroids, is about 20, with a total mass of order $10^{-9}$ $M_E$. This is much smaller than the mass of NEOs, so one can neglect the contribution of LP comets to $\mu$ for the inner planets. Similarly, the number of LP comets in the outer planetary zone ($r_o < 50$ AU) at any time that are potentially able to collide with giant planets or with KBOs is about $10^4$, with a total mass of order $10^{-7}$ $M_E$. This is much smaller than the mass of KBOs and Centaurs, so one can also neglect LP comets in estimating $\mu$ for the outer planets. For this reasons, I restrict the tally of potential colliders to those bodies that share the orbital zone of a target, defined such that the orbital periods of the colliding and target bodies differ by less than an order of magnitude.

## 5. Planetary Discriminants

Table 1 lists the estimated values of $\mu = M/m$ for the planets and for Ceres, Pluto and Eris. For a given target object, I take m to be the mass of *all* other members of the population that share its orbital zone. Thus I regard all NEOs as able to collide with any terrestrial planet, and all Centaurs as able to collide with any giant planet. This is a crude approximation, but the uncertainties in the total masses of the proximate populations are sufficient that a more detailed breakdown according to orbits seems unwarranted. I thus set m equal to the aggregate masses of the NEOs, of the MCOs and of the Centaurs, respectively, for the three inner planets, for Mars, and for the giant planets. Hence, $\mu = 1.7 \times 10^6$ M, $\mu = 5100$, and $\mu = 2000$ M for the three cases, respectively, where M is the planet mass relative to Earth.

An adjustment for the case of Neptune may be justified. Some 25% of the KBOs are Plutinos, which cross the orbit of Neptune with a 3:2 mean motion resonance. The resonance protects them from collision with the planet. Of the 783 KBOs with published orbital eccentricity, no more than 5 are non-resonant and free to collide with Neptune (Fig. 2). If these are a representative sample by mass of the entire KBO population, then the KBOs that can potentially collide with Neptune have a mass of (5/783) 0.03 $M_E = 2 \times 10^{-4}$ $M_E$. Adding this to the mass of the Centaurs decreases $\mu$ for Neptune by 40%.

Ceres, the largest asteroid, has a mass of $1.5 \times 10^{-4}$ $M_E$, about 1/4 the total mass of the asteroid belt. Most other asteroids can potentially collide with Ceres, which thus has $\mu = (1/4)/(3/4) = 1/3$. The asteroid belt could not produce a planet because its relative collision speeds, energized by primordial perturbations from planetesimals and Jupiter, result in net erosion rather than accretion for such small bodies (Bottke et al. 2005).

Pluto crosses the orbit of Neptune, but its 3:2 mean motion resonance with the planet shields it from a collision. However, most of the known KBOs cross the orbit of Pluto and can potentially collide with it. Pluto has a mass of 0.0022 $M_E$, about 7% of the Kuiper Belt's mass.

Figure 3 plots m versus M for heliocentric bodies, which fall into five classes according to the total mass m of the potential colliding population. Main belt asteroids and giant planets coincidentally share nearly the same value of m. The figure omits data points for asteroids and KBOs smaller than the first two of each class, but such objects would be plotted vertically below the points for Pallas and Pluto.

The planetary discriminant $\mu$ has a sharply bimodal distribution, with a gap of four orders of magnitude between the values for Mars and Ceres (Table 1). The solid lines in Figure 3 represent those limiting values, $\mu = 5100$ and 1/3, respectively. The eight planets Mercury through Neptune fall above the gap and everything else falls below it.

The magnitude of the gap is not simply due to the difference in mass between, say, Pluto and Mercury. The mass ratio for those bodies is only 25, but their ratio in $\mu$ is $10^6$ (Table 1). Rather, the gap reflects the fact that objects with high values of $\mu$ are fully accreted planets while those with low values were stranded in an arrested stage of development. This clear bifurcation again supports the designation of objects on one side of the gap as planets and those on the other side as non-planets. The latter include the asteroids, comets and KBOs.

The dashed diagonal line represents $\mu = 100$, which I suggest as a provisional boundary between planets and non-planets. It lies near the midpoint of the observed gap. The numerical choice is somewhat arbitrary but not critical. The gap is wide enough that a boundary anywhere between about 10 and 1000 would also be acceptable.

The observed planetary discriminant $\mu$ depends only on the mass of the target body and the present aggregate mass of potential colliding objects, while the theoretical scattering parameter $\Lambda$ depends only on the mass and orbital period of the target body. However, both parameters show a gap of four or five orders of magnitude between planets and non-planets.

## 6. Mass Limits and Accretion Hierarchy

The upper mass limit for a planet is often taken to be about 13 Jupiter masses ($M_J$), above which an interval of deuterium fusion occurs and the body is called a brown dwarf. Deuterium fusion is a relatively insignificant process compared to hydrogen fusion in stars (bodies with $M > 80\ M_J$).

Brown dwarfs may be as common as stars in the Galaxy, but they rarely occur as close companions of stars. Those that do are evidently products of disk accretion, like planets, and unlike the low-mass stellar secondaries, which belong to a different population (Mazeh et al. 2003). The rarity of close brown dwarf companions of stars, relative to the frequency of close planetary and stellar mass companions, is called the “brown dwarf desert” (Endl et al. 2004, Grether & Lineweaver 2006). Disk accretion rarely forms bodies exceeding 15 $M_J$, perhaps because most of the disk mass dissipates in $10^7$ years (Greaves 2005), leaving insufficient time for bodies to accrete more mass (Mohanty & Jayawardhana 2006).

The proposed definition of a planet as an end product of disk accretion around a primary star or substar removes the need for an upper mass limit. The relatively rare close companions with $M > 13\ M_J$ can be classified as planets.

The proposed definition of a planet also removes the need to assign a lower mass limit to distinguish planets from asteroids and comets, based for example on the hydrostatic criterion of spheroidal shape. A potato-shaped body would be classified as a planet if it dominated its orbital zone. Any mature system of planets will have a smallest member, but the mass of that member will depend on the history of the accretion process.

Objects with mass $< 13\ M_J$ are evidently formed by primary core collapse in molecular clouds as well as by secondary accretion from a disk (Tamura et al. 1998, Zapatero Osorio et al. 2000, Lucas &Roche 2000, Boss 2001). Such primary objects may have their own accretion disks and satellites. They are sometimes called “free-floating planets” to distinguish them from planets bound to stars or brown dwarfs. I provisionally refer to them as *sub brown dwarfs*.

This terminology allows us to sort all accreted objects by two criteria – mass range of the primary and level in an accretion hierarchy – as in Table 4.

1. “Primary” objects are stars, brown dwarfs or sub brown dwarfs, formed by core accretion from interstellar clouds.

2. “Secondary” objects are planets, defined as end products of disk accretion around a primary object.

3. “Tertiary” objects are either regular satellites (e.g., the Galilean moons), defined as end products of disk accretion around secondary objects (planets), or irregular satellites (captured bodies).

4. “Fourth-order” objects may have accreted in orbits around regular satellites in our solar system but we do not observe them. Tidal orbital evolution would have destroyed them early in the history of the solar system (Reid 1973). This happens because tides raised by planets on regular satellites rapidly despin them to synchronous rotation, and tides raised on a synchronous satellite by any substantial object in orbit around it would rapidly cause such an object to spiral into the satellite. Collisions with meteoroidal debris would destroy smaller fourth order satellites in less than the age of the solar system.

5. “Debris.” Other objects in planetary systems, including asteroids and comets, are the leftover and scattered debris of primary disk accretion. They need not all be small. The Oort Cloud may well contain objects as large as terrestrial planets (Stern 1991). However, such displaced planetary embryos would be very remote from their place of origin, would not dynamically control a well-defined volume of orbital space, and would not be classified as planets in the proposed system.

6. “Substars” include isolated brown dwarfs and sub brown dwarfs, both of which may have planets. They have two modes of origin: by primary core accretion in molecular clouds, and by secondary accretion in stable circumstellar disks followed by gravitational expulsion (Mohanty & Jayawardhana 2006). Substars of the second class may lose any planets in the process of expulsion.

7. “Satellites of substars.” Bodies formed as end products of disk accretion in orbit around primary brown dwarfs are here called planets. Many brown dwarfs have disks (Jayawardhana et al. 2003), which suggests that planets around such objects are common. Some sub brown dwarfs ($< 13\ M_J$) also appear to have disks (Luhman et al. 2005, Allers et al. 2006), suggesting that they too may have their own satellites. Such secondaries may form in the same way as the regular satellites of giant planets.

8. “Rogue planet” is a term applied to a low mass object ($< 13\ M_J$) accreted in a disk and ejected to interstellar space by gravitational perturbations. Simulations of planetary formation and migration suggest that such escaped planets are very numerous (Moorhead & Adams 2005). However, they would be difficult to detect and, if detected, might be indistinguishable from sub brown dwarfs, which accreted as primary objects.

## 7. Exoplanetary Systems

For our proposed definition of a planet to have general validity, exoplanets should also have non-overlapping orbits, unless shielded from collisions by a mean motion resonance. Among 164 exoplanetary systems catalogued by Schneider (2006), 20 are known to possess more than one planet. Figure 4 shows the distance ranges (from pericenter to apocenter) of the known secondaries in those systems, together with the inner five planets of our own solar system.

Most of these exoplanetary systems have non-intersecting orbits, with three exceptions. In HD 128311, the two orbits appear to approach each other within 0.1 AU, and the uncertain eccentricities allow that the orbits may actually cross. However, these planets appear to share a 2:1 mean motion resonance (Vogt et al. 2005, Sandor & Kley 2006), which would prevent them from colliding. In HD 82943, the two planets have overlapping orbits, but they are probably also in a 2:1 resonance (Ferraz-Mello et al. 2005).

For the system HD 160691, the best-fit solution to the Doppler observations (McCarthy et al. 2005) allows the orbits of the two outer planets to overlap, in which case the system would become unstable in < 20,000 years (Gozdziewski et al. 2005). However, uncertainties in the orbit of the outer planet allow for the possibility that the orbits do not overlap and/or that a mean motion resonance stabilizes the system (Bois et al. 2003, Gozdziewski et al. 2005).

Orbital migration of growing planets interacting with the disk planetesimals may establish mean motion resonances between planets, which would allow them to survive on intersecting orbits without collisions.

All known exoplanets of main sequence stars fall well above the gap in Figure 1 and would be classified as planets by the criterion of dynamical dominance.

## 8. Conclusions

I propose to define a planet as an end product of secondary accretion in a disk around a primary star or substar. Planets in this sense occur only in highly evolved (old) systems, which have reached the final cleanup phase of accretion, with the major bodies in stable non-intersecting or resonant orbits. The definition derives solely from the basic physics of the formation of planetary systems.

Planets are the solitary bodies that prevail in the creative-destructive evolution of a disk, and are dynamically distinct from the populations of leftover debris -- mainly asteroids and comets. The difference between planets and non-planets is quantifiable, both theoretically and observationally. All planets in our solar system are sufficiently massive to have scattered most of the remaining planetesimals out of their orbital zones in less than a Hubble time. Today these planets dominate the residual mass in their orbital zones by many orders of magnitude.

The proposed definition of a planet does not depend on upper or lower mass limits, the deuterium fusion threshold, or approximation to hydrostatic equilibrium. Any body that orbits a star or substar and contains more than about 100 times the mass of all other bodies in its orbital zone is a planet.

The end product of secondary disk accretion is a small number of relatively large bodies (planets) in either non-intersecting or resonant orbits, which prevent collisions between them. The proposed definition for a planet is consistent with what we know about extrasolar planetary systems.

A prominent objection to any definition of this kind is the contention that natural objects should be defined only by their intrinsic properties (mass, shape, composition, etc.) and not by their dynamical context. But why is dynamical context any less relevant? We refer to objects that orbit planets as “moons”, although two of them are larger than the planet Mercury, and most of them (called irregular satellites) appear to be captured asteroids and comets. We distinguish meteoroids orbiting the Sun from meteorites lying on the ground. The definition proposed here distinguishes planets, which dynamically dominate a well-defined volume of orbital space, from asteroids, KBOs and ejected planetary embryos, which do not.

One can probably devise hypothetical cases that resist classification by the definition proposed here, but “pathological” cases occur in most sciences, and do not invalidate our understanding of basic relationships.

The historical definition of “nine planets” no doubt retains a strong sentimental attraction (Basri & Brown 2006). However, *ad hoc* definitions devised to retain Pluto as a planet tend to conceal from the public the paradigm shift that has occurred since the 1990s in our understanding of the architecture of the solar system, in relation to its origin in a solar nebula. To be useful, a scientific definition should be derived from, and draw attention to, the basic principles.

**Acknowledgments:** I am grateful to Neil deGrasse Tyson for encouraging me to write this paper and for stimulating discussions of these ideas. I also thank Michael Shara, Michal Simon, Alan Stern and Vincenzo Zappalà for useful comments.

TABLES

Table 1
Planetary discriminants

| **Body** | **Mass ($M_E$)** | **Λ** | **μ** |
|---|---|---|---|
| Mercury | 0.055 | $1.9 \times 10^3$ | $9.1 \times 10^4$ |
| Venus | 0.815 | $1.7 \times 10^5$ | $1.35 \times 10^6$ |
| Earth | 1.000 | $1.5 \times 10^5$ | $1.7 \times 10^6$ |
| Mars | 0.107 | $9.3 \times 10^2$ | $5.1 \times 10^3$ |
| Ceres | 0.00015 | 0.0013 | 0.33 |
| Jupiter | 317.7 | $1.3 \times 10^9$ | $6.25 \times 10^5$ |
| Saturn | 95.2 | $4.7 \times 10^7$ | $1.9 \times 10^5$ |
| Uranus | 14.5 | $3.8 \times 10^5$ | $2.9 \times 10^4$ |
| Neptune | 17.1 | $2.7 \times 10^5$ | $2.4 \times 10^4$ |
| Pluto | 0.0022 | 0.003 | 0.07 |
| Eris | 0.0028 | 0.002 | 0.10 |

Table 2
Primary reservoirs of colliding objects

| **Reservoir** | **a (AU)** | **Number** | **Mass ($M_E$)** | **Ref. for Mass** |
|---|---|---|---|---|
| Asteroid Belt | 2 to 3.5 | $1.2 \times 10^6$ | 0.0006 | Krasinski et al. 2002 |
| Kuiper Belt | 35 to 56 | $6 \times 10^8$ | 0.03 | Bernstein et al. 2004 |
| Oort Cloud | $10^3$ to $10^5$ | $5 \times 10^{11}$ | 2 to 40 | Francis 2005 |

Table 3
Proximate populations of colliding objects

| **Population** | **q (AU)** | **Number** | **Mass ($M_E$)** | **Ref. for Number** |
|---|---|---|---|---|
| NEOs | < 1.3 | $10^3$ | $6 \times 10^{-7}$ | Stuart & Binzel 2004 |
| MCOs | 1.3 – 1.7 | $3.5 \times 10^4$ | $2 \times 10^{-5}$ | Michel et al. 2000 |
| Centaurs | > 5 | $10^7$ | $5 \times 10^{-4}$ | Shepard et al. 2000 |
| SP Comets | < 5 | $10^3$ | $5 \times 10^{-8}$ | Levison et al. 2002 |
| LP Comets* | < 50 | $10^4$ | $5 \times 10^{-7}$ | This paper |

*For LP comets, the number and mass are for comets presently at r < 50 AU.

Table 4
Classification of accreted objects

| Primary | Secondary | Tertiary |
| --- | --- | --- |
| star (> 80 $M_J$) | planet | moon |
| brown dwarf (13 – 80 $M_J$) | planet | moon |
| sub brown dwarf (< 13 $M_J$) | planet | -- |

FIGURES

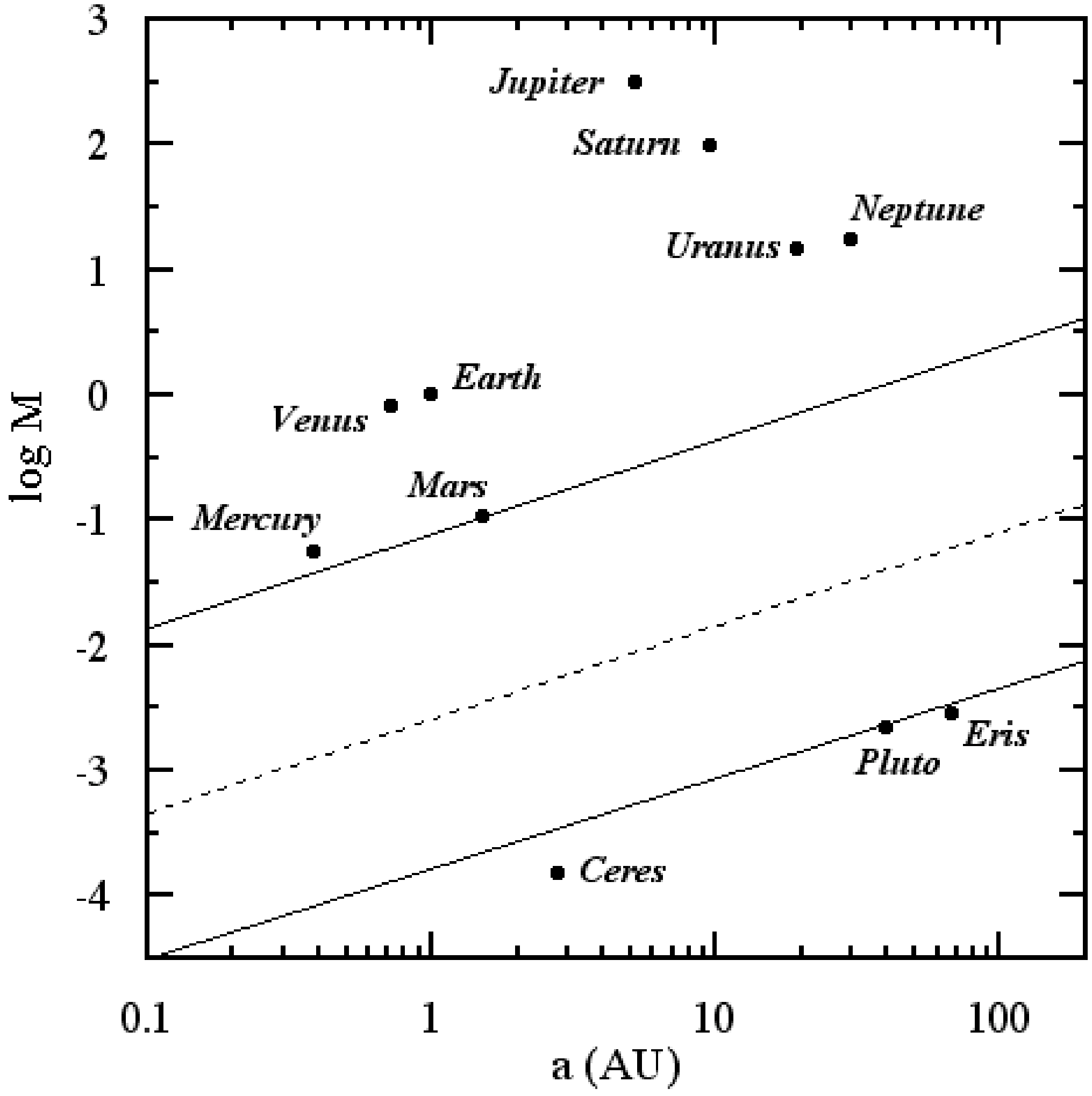

Fig. 1. Plot of mass M (in Earth masses) versus semi-major axis a for heliocentric bodies. The solid lines bound the gap in observed values of the scattering parameter $\Lambda = kM^2/P$, where P is the orbital period. The upper and lower lines represent $\Lambda$ values corresponding to those of Mars and Pluto, respectively. Any body above the dashed line, which represents $\Lambda = 1$, will scatter a significant fraction of planetesimals out of its orbital zone within a Hubble time.

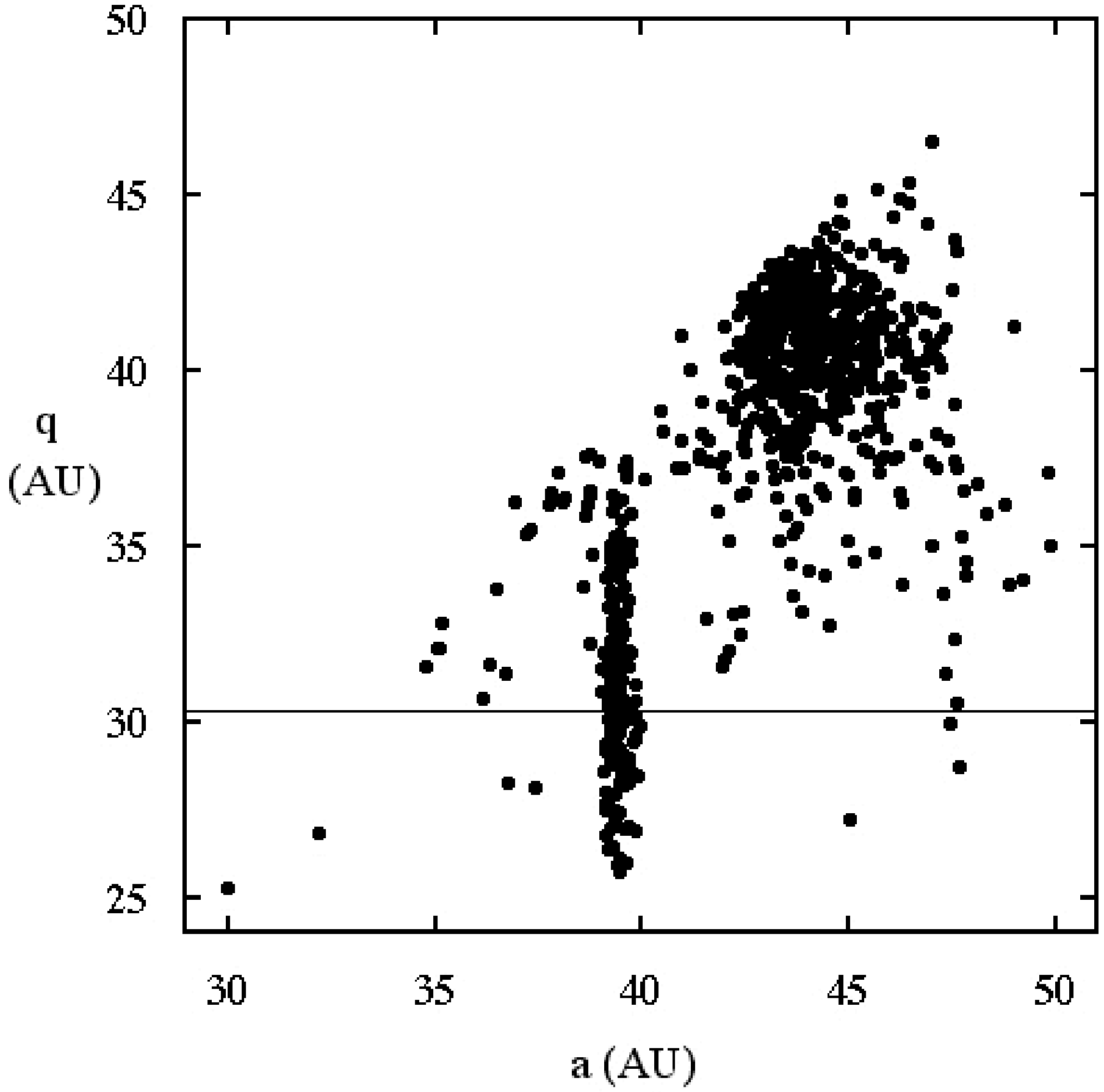

Fig. 2. Perihelion versus semi-major axis of 783 KBOs with well-determined orbits. Mean motion resonances with Neptune of order 3:2 and 2:1 are at a = 39.2 and 47.5 AU, respectively. The line marks q at the aphelion distance of Neptune.

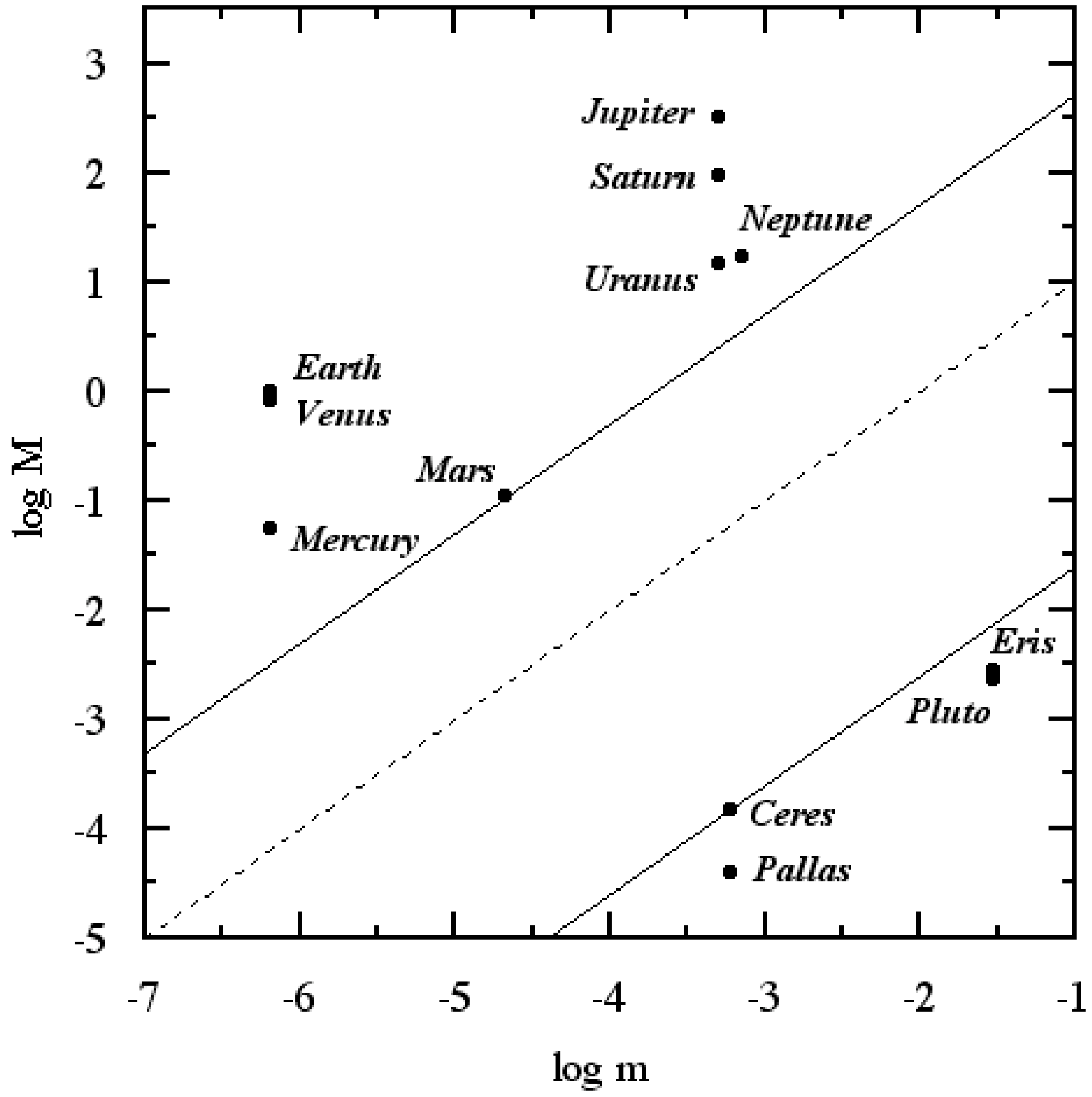

Fig. 3. Plot of mass M of a body versus the aggregate mass m in its orbital zone. The solid lines bound the observed gap in the ratio $\mu = M/m$, where $\mu = 5100$ and 1/3 are the ratios for Mars and Ceres, respectively. The dashed line represents $\mu = 100$. M and m are in Earth masses.

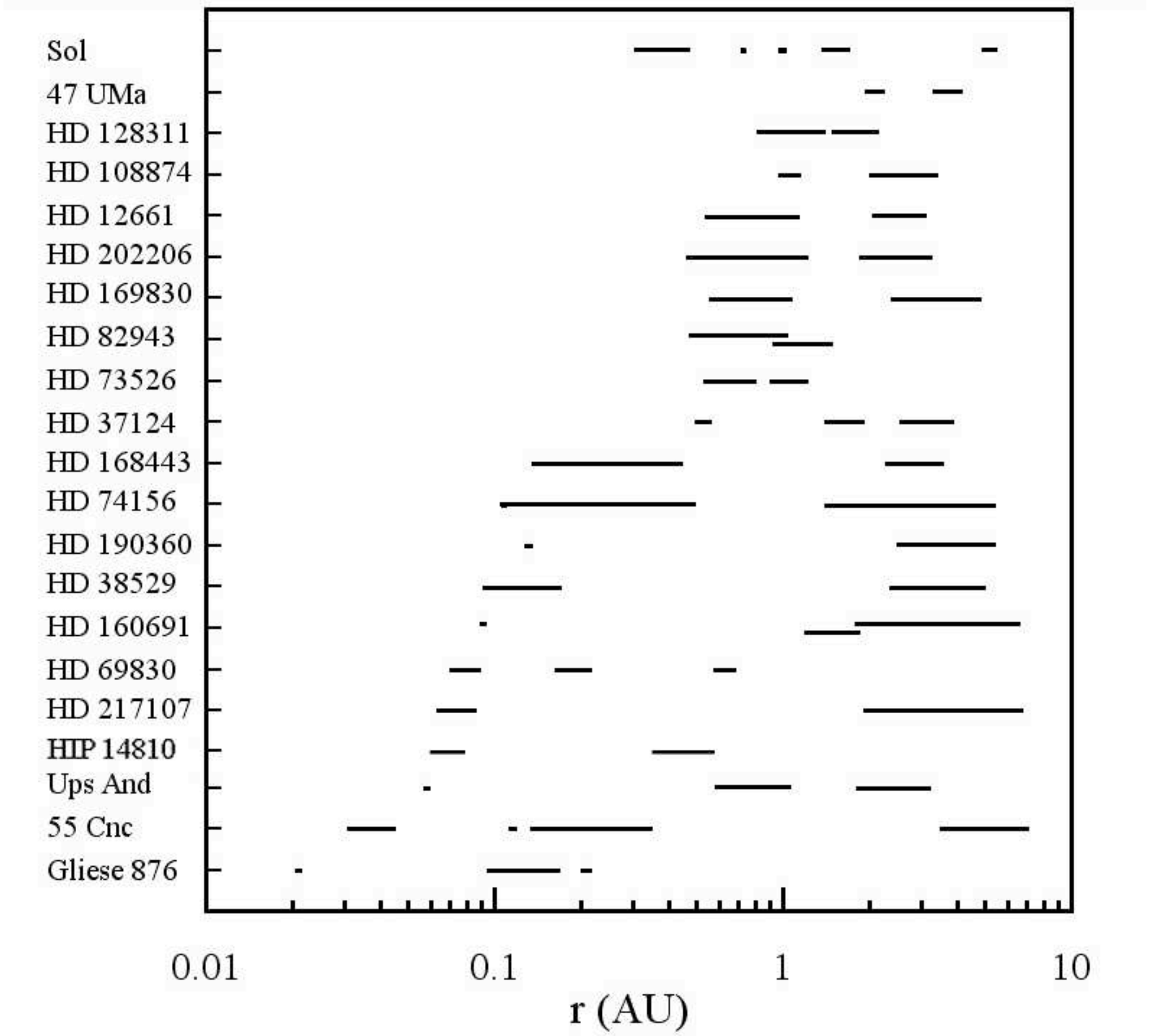

Fig. 4. Plot of 21 multi-planet systems including our solar system (Mercury to Jupiter). Line segments extend from pericenter to apocenter of each planet. Data from Schneider (2006).